\documentclass[twocolumn]{aastex6}

\usepackage{amsmath}
\usepackage{epstopdf}
\newcommand{\sunrise}{\textsc{Sunrise}}
\newcommand{\carcsec}{$\mbox{.\hspace{-0.5ex}}^{\prime\prime}$}

\begin{document}

\title{Moving Magnetic Features around a Pore}

\author{A. J. Kaithakkal\altaffilmark{1}, 
T. L. Riethm\"uller\altaffilmark{1},
S. K. Solanki\altaffilmark{1,2},
A. Lagg\altaffilmark{1},
P. Barthol\altaffilmark{1},
A. Gandorfer\altaffilmark{1},
L. Gizon\altaffilmark{1},
J. Hirzberger\altaffilmark{1},
M. vanNoort\altaffilmark{1},
J. Blanco Rodr\'iguez\altaffilmark{3},
J. C. Del Toro Iniesta\altaffilmark{4},
D. Orozco Su\'arez\altaffilmark{4},
W. Schmidt\altaffilmark{5},
V. Mart\'inez Pillet\altaffilmark{6},
\& M. Kn\"olker\altaffilmark{7}
}
\altaffiltext{1}{Max Planck Institute for Solar System Research,
Justus-von-Liebig-Weg 3, G\"ottingen 37077, Germany; anjali@mps.mpg.de}
\altaffiltext{2}{ School of Space Research, Kyung Hee University, Yongin, Gyeonggi, 446-701, Korea}
\altaffiltext{3}{Grupo de Astronom\'ia y Ciencias del Espacio, Universidad de Valencia, 46980 Paterna, Valencia, Spain}
\altaffiltext{4}{Instituto de Astrof\'isica de Andaluc\'ia (CSIC), Apartado de Correos 3004, 18080 Granada, Spain}
\altaffiltext{5}{Kiepenheuer-Institut f\"ur Sonnenphysik, Sch\"oneckstr. 6, 79104 Freiburg, Germany}
\altaffiltext{6}{National Solar Observatory, 3665 Discovery Drive, Boulder, CO 80303, USA}
\altaffiltext{7}{High Altitude Observatory, National Center for Atmospheric Research,\footnote{The National Center for Atmospheric Research is sponsored by the National Science Foundation.} P.O. Box 3000, Boulder, CO 80307-3000, USA}

\begin{abstract}
Spectropolarimetric observations from \sunrise{}/IMaX obtained in June 2013 are used for a statistical analysis to determine the physical properties of moving magnetic features (MMFs) observed near a pore. MMFs of the same and opposite polarity with respect to the pore are found to stream from its border at an average speed of 1.3 km s$^{-1}$ and 1.2 km s$^{-1}$ respectively, with mainly same-polarity MMFs found further away from the pore. MMFs of both polarities are found to harbor rather weak, inclined magnetic fields. Opposite-polarity MMFs are blue-shifted, while same-polarity MMFs do not show any preference for up- or downflows. Most of the MMFs are found to be of sub-arcsecond size and carry a mean flux of $\sim$ 1.2$\times 10^{17}$ Mx.
\end{abstract}
 

\section{Introduction} \label{sec:intro}
The presence of moving magnetic features (MMFs) around sunspots was first noted by \cite{sh}, who found that there are small-scale bright features moving out radially from mature sunspots with speeds of about 1 km s$^{-1}$. Later \cite{har} found that these moving bright features are magnetic and named them MMFs. The majority of the studies on MMFs in the literature was focused on proving the connection between penumbral magnetic fields and the origin of MMFs \citep[and references therein]{zh1,sainz,sol,zh2,kuboa,kubob} under the assumption that a penumbra is necessary for MMFs to be produced. However, \cite{har} showed that a penumbra is not essential for the presence of MMFs. Later studies have confirmed that MMFs are observed around sunspots without a penumbra, or more precisely, pores \citep{zu,ver,ser}.

Studies have further shown that MMFs appear as either unipolar magnetic features or bipolar feature pairs. The unipolar MMFs can have either the same or opposite polarity with respect to the parent spot \citep{sh1}. Their typical size is below 2$\arcsec$ and they exhibit a broad range of horizontal velocity from 0.1 to 1.5 km s$^{-1}$ \citep{har,bri,zh1}. Unipolar features with opposite polarity to the spot are reported to have higher speeds than features that are parts of bipolar pairs and unipolar features with the same polarity as the spot \citep[for a review on MMFs see][]{hag}. 

A recent study by \cite{ser}, using data from IBIS at NSO/DST, which investigated 6 unipolar MMFs around a pore, out of which 3 were of opposite polarity to the pore, confirmed that opposite-polarity MMFs move faster than same-polarity MMFs. Furthermore, MMFs of opposite polarity were shown to be associated with upflows while those of the same-polarity as the pore were found to be associated with downflows. The authors also pointed out that the characteristics of MMFs observed around a pore are in general consistent with those obtained for MMFs around sunspots. 

Despite the existing studies on MMFs it is not yet clear to what extent the origin of MMFs is related to the presence of a penumbra or not. Also, there are very few studies that provide information on the physical properties of the MMFs. In this paper, we present a statistical analysis of physical properties of MMFs observed near a pore. We used high spatial resolution ($\sim$ 0\carcsec{}15) and high cadence (36.5 s) spectropolarimetric data obtained during the second flight of the balloon-borne observatory \sunrise{} \citep{bart,gan,berk,samia,samib} for our analysis.

\section{OBSERVATION AND DATA ANALYSIS} \label{sec:obsv}
\begin{figure*}[htbp]
  \includegraphics[trim=2 2 2 2,clip,width=1.0\textwidth]{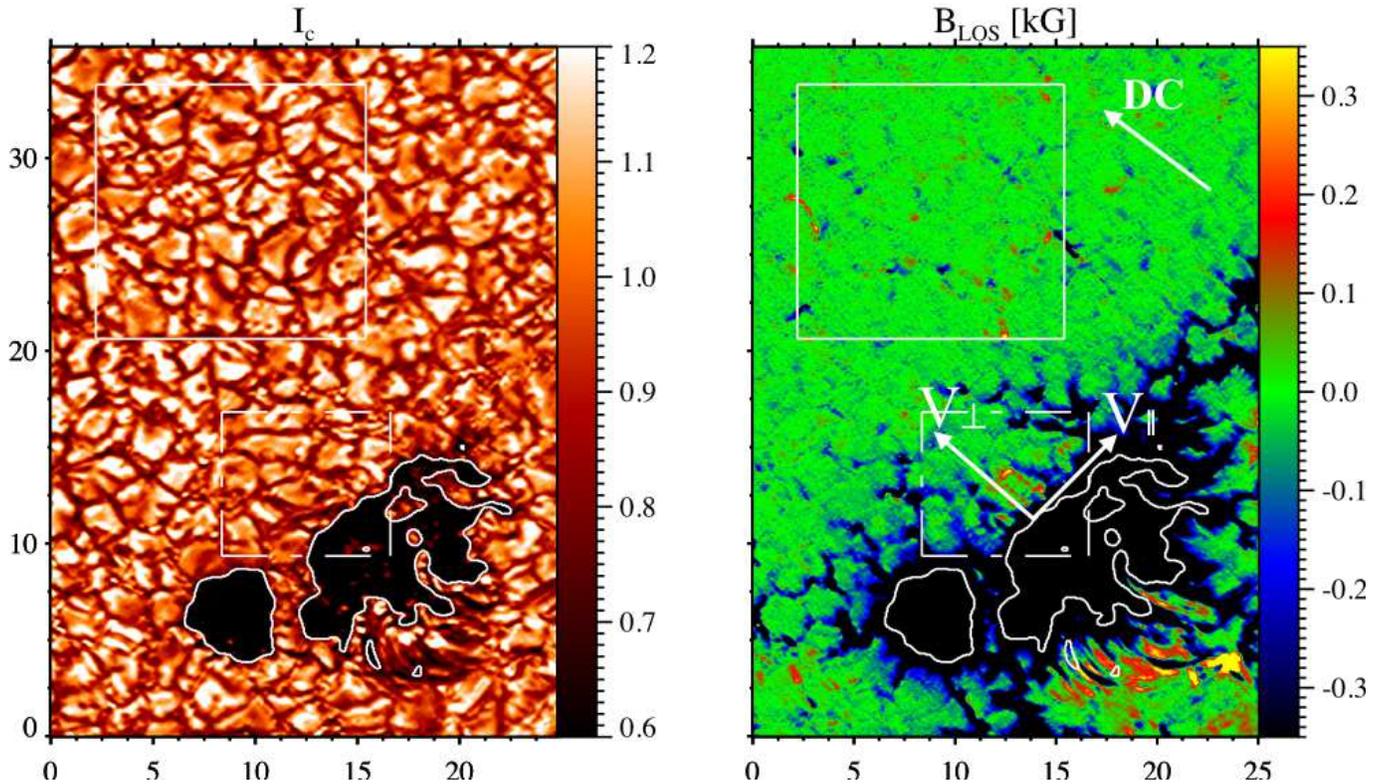}
  \caption{Left: Normalized continuum intensity map of the first frame of the analyzed time series. Right: Corresponding B$_{LOS}$ map saturated to $\pm$ 350 G. $V_\perp$ and $V_\parallel$ represent components perpendicular and parallel to the pore boundary respectively, of the proper motion velocity of the MMFs. The white box on the upper left corner indicates the quite-Sun region and the white dashed box outlines our region of interest (ROI). The white contours outlining the edges of the pores represent the continuum intensity level of 0.69$I_{c,QS}$, where $I_{c,QS}$ is the continuum intensity averaged over the quite-Sun region. DC corresponds to disk center and the arrow point towards the approximate disc center direction. The axes are in arcsec.\label{fig:a}}
\end{figure*}

The field of view (FOV) of IMaX/ \sunrise{} II observations carried out on 12 June 2013, 23:39 - 23:55 UT, covered a large pore which is part of the active region AR 11768 ($\mu$ = 0.93; see Figure \ref{fig:a}). IMaX recorded the full Stokes vector over the \ion{Fe}{1} 5250.2 {\AA} (Land\'e factor, g = 3) spectral line in the V8-4 mode \citep[for details see ][]{pillet}, i.e. at seven wavelength positions ($\lambda = \pm$120, $\pm$80, $\pm$40 m\AA, and the line center) within the line and one in the nearby continuum (+227 m\AA). The FOV of the images is 51\arcsec x 51$\arcsec$ with a scale of 0\carcsec{}0545 per pixel.

The images were corrected for dark current, flat field, fringes, polarization crosstalk and stray light. Afterwards the data were reconstructed with the help of a point spread function retrieved from inflight phase-diversity measurements. The reconstructed IMaX data have a spatial resolution of 0\carcsec15 -- 0\carcsec18. The noise ($\sigma$) in Stokes $V$ is around 7$\times$10$^{-3}I_c$, where $I_c$ is the continuum intensity. The Stokes vectors were inverted using the classical SPINOR code \citep{fru} that uses the STOPRO routines \citep{samic}, assuming a single component model atmosphere. The inversion was set to return the temperature at three optical depths, and height independent magnetic field parameters (B, $\gamma$ and $\phi$), line-of-sight (LOS) velocity ($v_{LOS}$) and microturbulent velocity. $v_{LOS}$ is corrected for the blueshift across the FOV due to the collimated setup of the Fabry-P$\acute{e}$rot etalon. More details on the data reduction and inversions are provided by \cite{samib} and references therein. The signature of p-modes was removed from the continuum and $v_{LOS}$ maps using a subsonic filter \citep{title} with a cut-off phase velocity of 4 km s$^{-1}$. The zero (reference) point for $v_{LOS}$ maps is defined as the spatio-temporal average value within the quiet-Sun region in the FOV (region within the solid white box in Figure \ref{fig:a}) and is subtracted from all the frames. The continuum intensity maps are normalized with respect to the mean quiet-Sun value. Image sequences of all the relevant parameters were corrected for rotation of the FOV caused by the alt-azimuth mounting of the \sunrise{} telescope, and then aligned using a spatial cross-correlation technique. We then chose a small area (denoted by the dashed white square in Figure \ref{fig:a}) with a size of 8\carcsec{}3 $\times$ 7\carcsec{}5 as our region-of-interest (ROI). 

\section{Identification and Tracking of Moving Magnetic Features}
MMFs were identified by applying a modified version of the multilevel tracking (MLT) algorithm of \cite{bov} to maps of the LOS component of the magnetic field vector, $B_{LOS}$. MLT uses multiple threshold levels with decreasing values of $B_{LOS}$. The area of the feature grows with lower thresholds. If the separation between two given features identified by two consecutive thresholds is less than 3 pixels they are joined into one single feature. The process of identifying and separating the features based on the thresholds was repeated until the lowest threshold level was reached. After a number of trials, we chose 40 thresholds ranging from 500 G to 40 G (the latter is the 3$\sigma$ value obtained from the $B_{LOS}$ maps). The code returned spatial locations of the features, each of which was tagged with a unique identification number based on its polarity. We then calculated the magnetic centroid of each feature, which is defined as the average of the position $(x_i, y_i)$ weighted by $B_{LOS}(x_i,y_i)$. We selected those features which 1) have a minimum size of 5 pixels (0.1 Mm in diameter); 2) have a per pixel $B_{LOS}$ value greater than 40 G; and 3) are within 3 Mm distance from the visible pore boundary. The visible boundary of the pore is defined as where $I_{c}$ is 69$\%$ of the quiet-Sun value. It should be noted that the magnetic boundary of the pore is larger than its visible boundary. For simplicity, we henceforth call visible pore boundary/border as pore border/boundary.

\begin{figure*}[htbp]
\includegraphics[trim= 18 346 170 295,clip,width=1.0\textwidth]{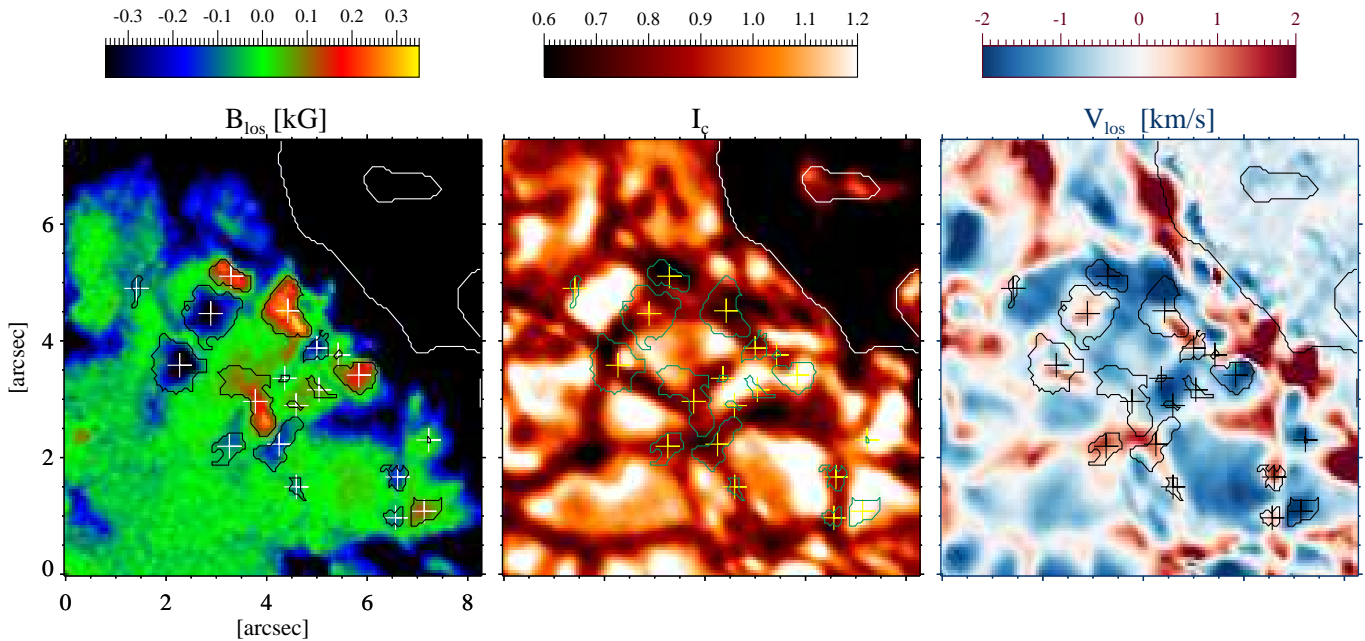}
\caption{Left to right: $B_{LOS}$, $I_c$/$<I_{c,QS}>$, and v$_{LOS}$ maps of our ROI taken at 3 minutes after the start of the observation (please note that the image is reversed in the vertical direction with respect to the image in Figure \ref{fig:a}). The white contour outline the pore border and the black contours (green in central panel) enclose the chosen features. The + signs represent the magnetic centroid positions of the MMFs. (A movie is available in the online journal).\label{fig:b}}
\end{figure*}

MMFs were tracked using spatial overlap \citep{iida} over consecutive $B_{LOS}$ maps assuming a maximum advection velocity of 4 km s$^{-1}$ (corresponding to a displacement of 4 IMaX pixels within 36.5 sec). If there were multiple features in the successive frame that spatially overlap with a given feature from the previous frame, the one with the minimum difference in flux with respect to the given feature was chosen. From the MMFs selected from the MLT method only those with a minimum life time of four frames (2.4 minutes) were chosen. In total, we selected 88 MMFs which satisfy all the above conditions. We then obtained values of all relevant parameters (such as field strength ($B$), inclination ($\gamma$), normalized continuum intensity, etc.) at the magnetic centroid position. Further we calculated the perpendicular ($V_\perp$) and parallel ($V_\parallel$) components (the direction being defined with respect to the pore border) of the proper motion velocity of the MMFs (see right panel of Figure \ref{fig:a}). Figure \ref{fig:b} shows $B_{LOS}$ ($= Bcos\gamma$), $I_c/<I_{c,QS}>$, and $v_{LOS}$ maps, taken at 3 minutes after the start of the observation, of our ROI with contours outlining MMFs we selected. 

\section{Results} \label{sec:results}
We identified 44 opposite and 44 same-polarity MMFs. From a visual inspection of the transverse field maps in the region, we find horizontal fields pointing from some of the opposite-polarity MMFs to the same-polarity MMFs, suggesting that the two form an MMF pair. As these MMFs are surrounded by many other neighboring MMFs it is difficult to follow them as pairs. Hence, we treated them as individual MMFs to obtain their physical properties. Nevertheless, the possible connection of these MMF-pairs would be worth looking into in more detail as part of future studies.

\begin{figure*}[ht]
  $\begin{array}{rl}
  {\includegraphics[width=0.5\textwidth]{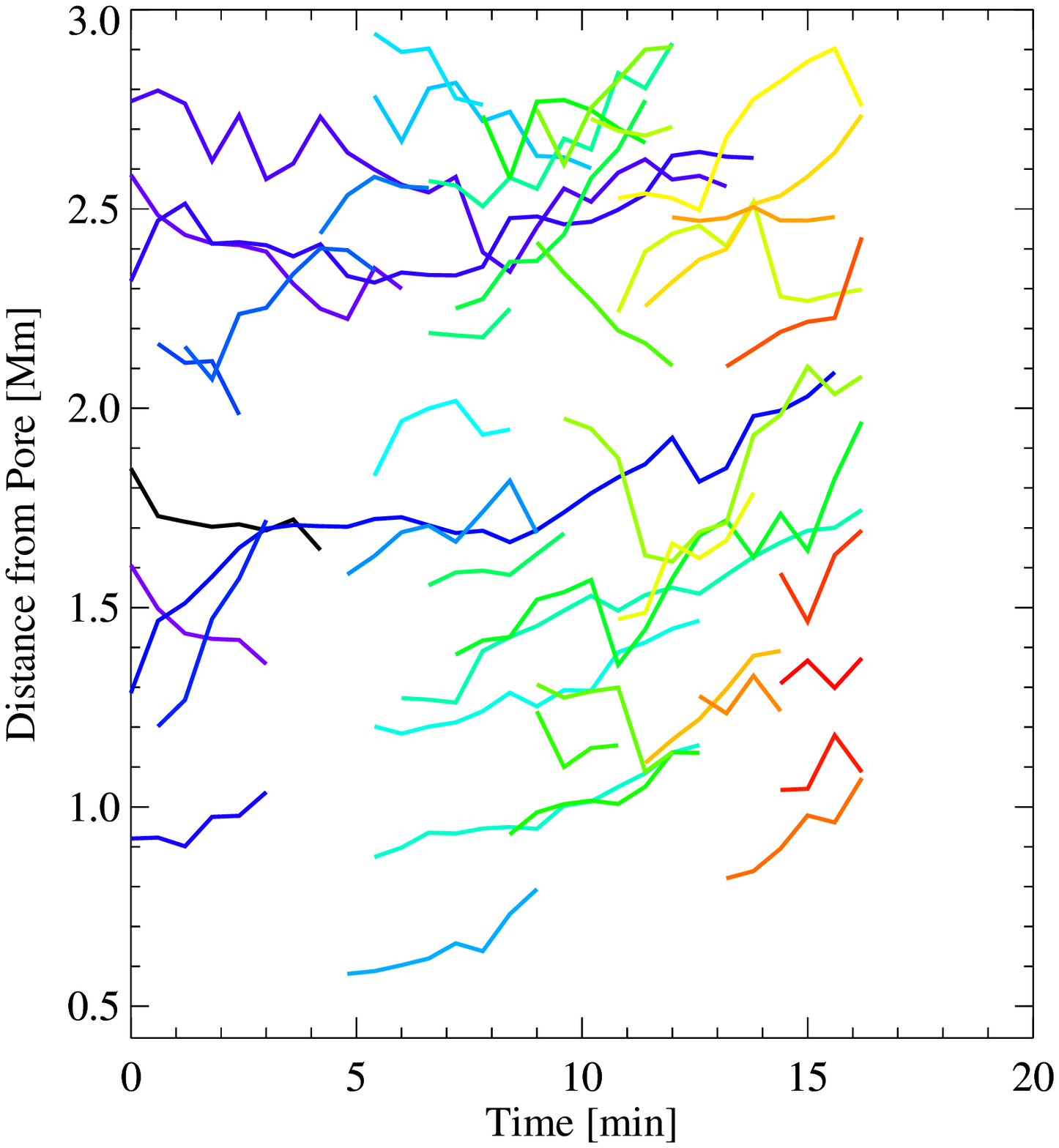}}&
   {\includegraphics[width=0.5\textwidth]{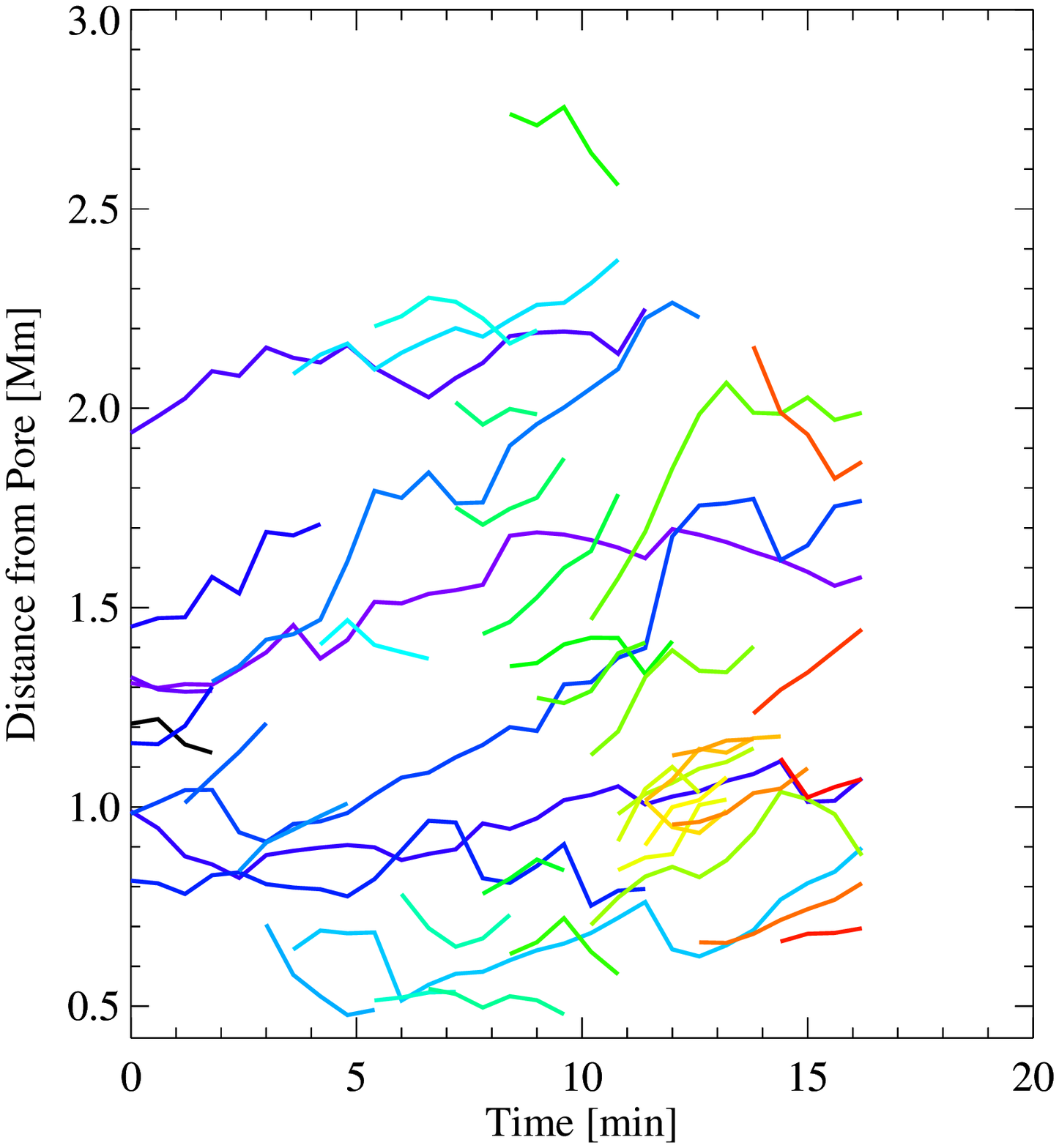}}
\end{array}$
  \caption{Left: Variation of the minimum distance from the pore border with time for the same-polarity MMFs. Right: Same as the left panel but for opposite-polarity MMFs. Different colors represent tracks of different MMFs.\label{fig:c}}
\end{figure*}

Figure \ref{fig:c} presents the variation of the minimum distance of the magnetic centroid of the MMFs from the pore border with time. It shows that for the majority of the MMFs of both polarities, the minimum distance from the boundary of the pore increases with time. When first identified, a few of the MMFs were  as close to the pore border as 0.6 Mm. We find that closer to the pore MMFs of opposite polarity dominate. The number of opposite-polarity MMFs within a distance of 1.5 Mm from the pore boundary is twice that of the same-polarity MMFs. It is possible that the strong magnetic field of the pore is masking the same-polarity MMFs closer to the pore, thereby rendering them less well visible. As described by \cite{ser} same-polarity MMFs could be detected when they are not embedded in the ambient field of the pore.

The angular distribution of the path followed by the MMFs shows a broad peak in the range [150$^\circ$ -- 180$^\circ$] where 180$^\circ$ represents motion in radial outward direction. Thus, although the MMFs are buffeted by the granulation their motion is not a pure random walk. We find that some of these MMFs move in a tangential direction to the pore boundary, while some others exhibit a random direction of motion for some period during their observed lifetime. However, most of the MMFs (72$\%$) eventually move away from the pore border. As we do not find any difference in the physical properties between MMFs based on the path they follow, we didn't differentiate between them when presenting our results in the following.

For each MMF, relevant physical parameters at the magnetic centroid position were determined and then those values were averaged over the period for which the MMF was observed. However, the whole feature was considered for the calculation of magnetic flux and area. We estimated the magnetic flux of each MMF as $\phi = B_{LOS}A$, where $B_{LOS}$ represents the LOS component of the magnetic field vector, and A the area of the resolution element. 

We averaged magnetic flux and area over the time for which each MMF could be tracked. For MMFs of both polarities, the area distribution is in the range 0.01-0.5 Mm$^2$ (i.e., the smallest ones are close to the spatial resolution of the \sunrise{}/IMaX observations) and the mean area is about 0.1 Mm$^2$. The magnetic flux distribution of the opposite-polarity MMFs lies in the range 7.7$\times10^{15}$ -- 9.5$\times10^{17}$ Mx, with the mean flux being 1.16$\times 10^{17}$ Mx. For the same-polarity MMFs the magnetic flux distribution ranges from 5.8$\times10^{15}$ to 6.6$\times10^{17}$ Mx, with a mean value of 1.19$\times 10^{17}$ Mx. 



Distributions of B, $\gamma$, $v_{LOS}$, $V_\perp$ and $I_c/<I_{c,QS}>$ averaged over the observed period of the MMFs corresponding to the magnetic centroid position are shown in Figure \ref{fig:d}. The details of the same are presented below:

\begin{figure*}[ht]
$\begin{array}{rl}
  {\includegraphics[width=0.4\textwidth]{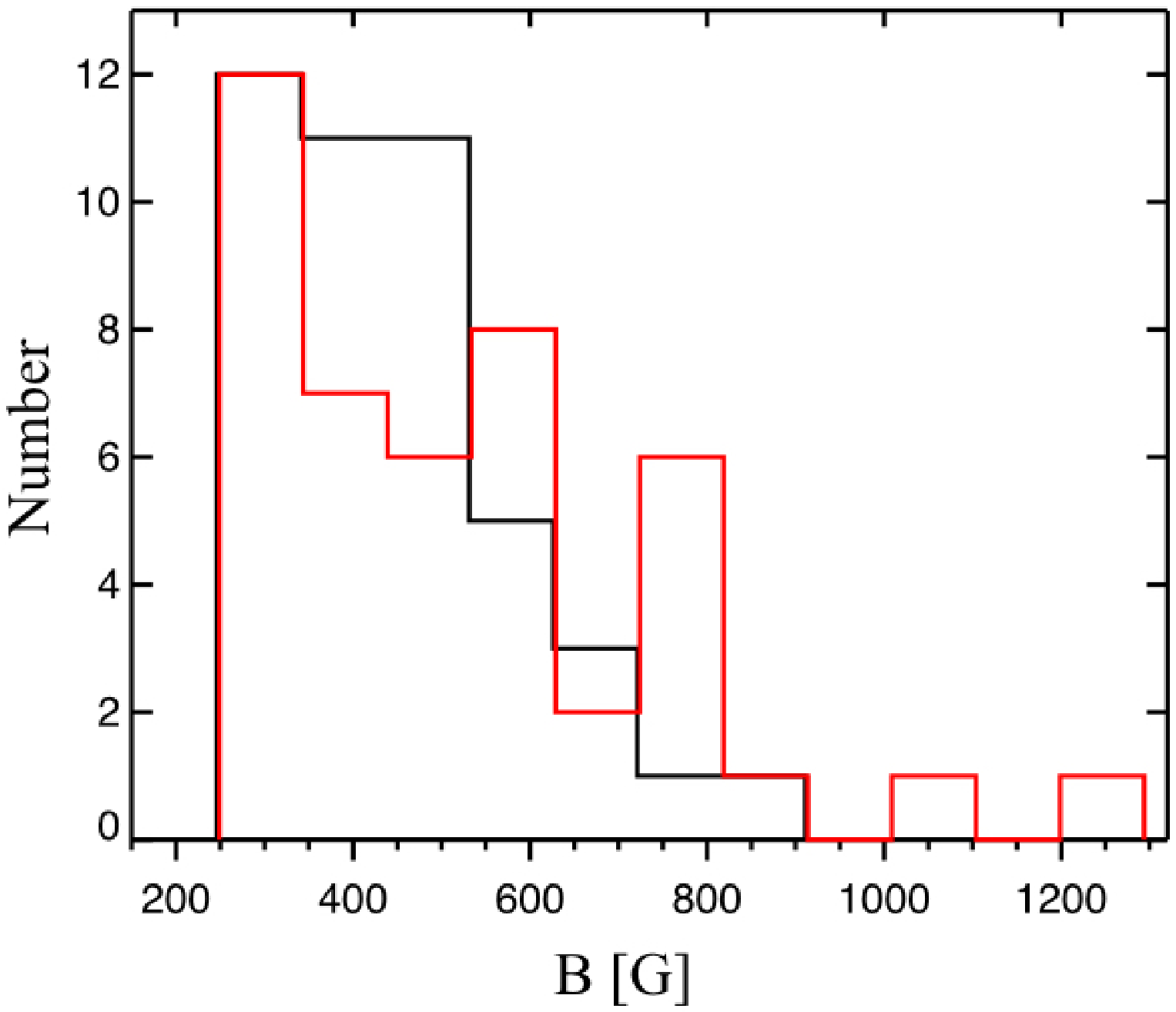}}(a)&
  {\includegraphics[width=0.4\textwidth]{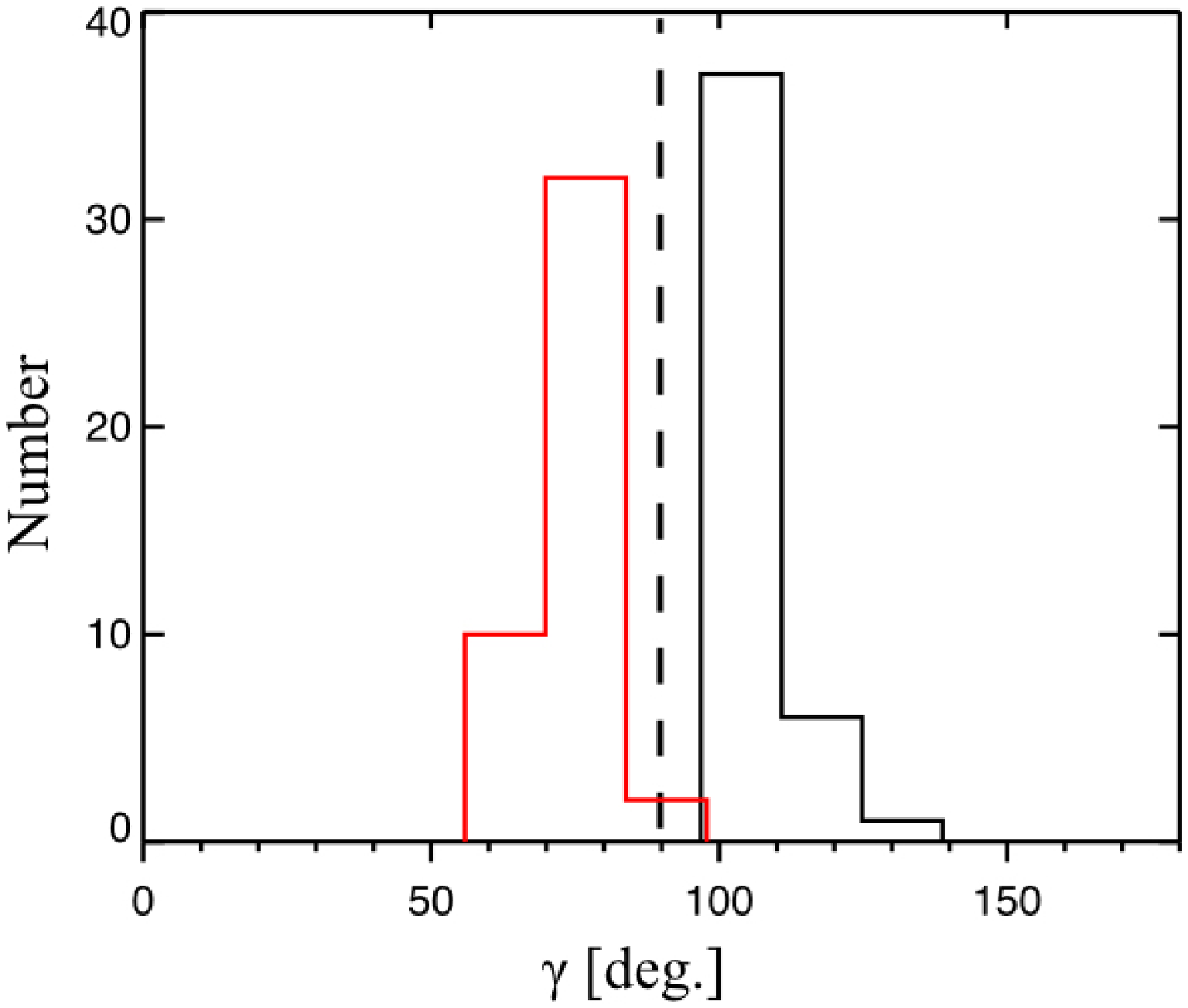}}(b)\\
   
    {\includegraphics[width=0.4\textwidth]{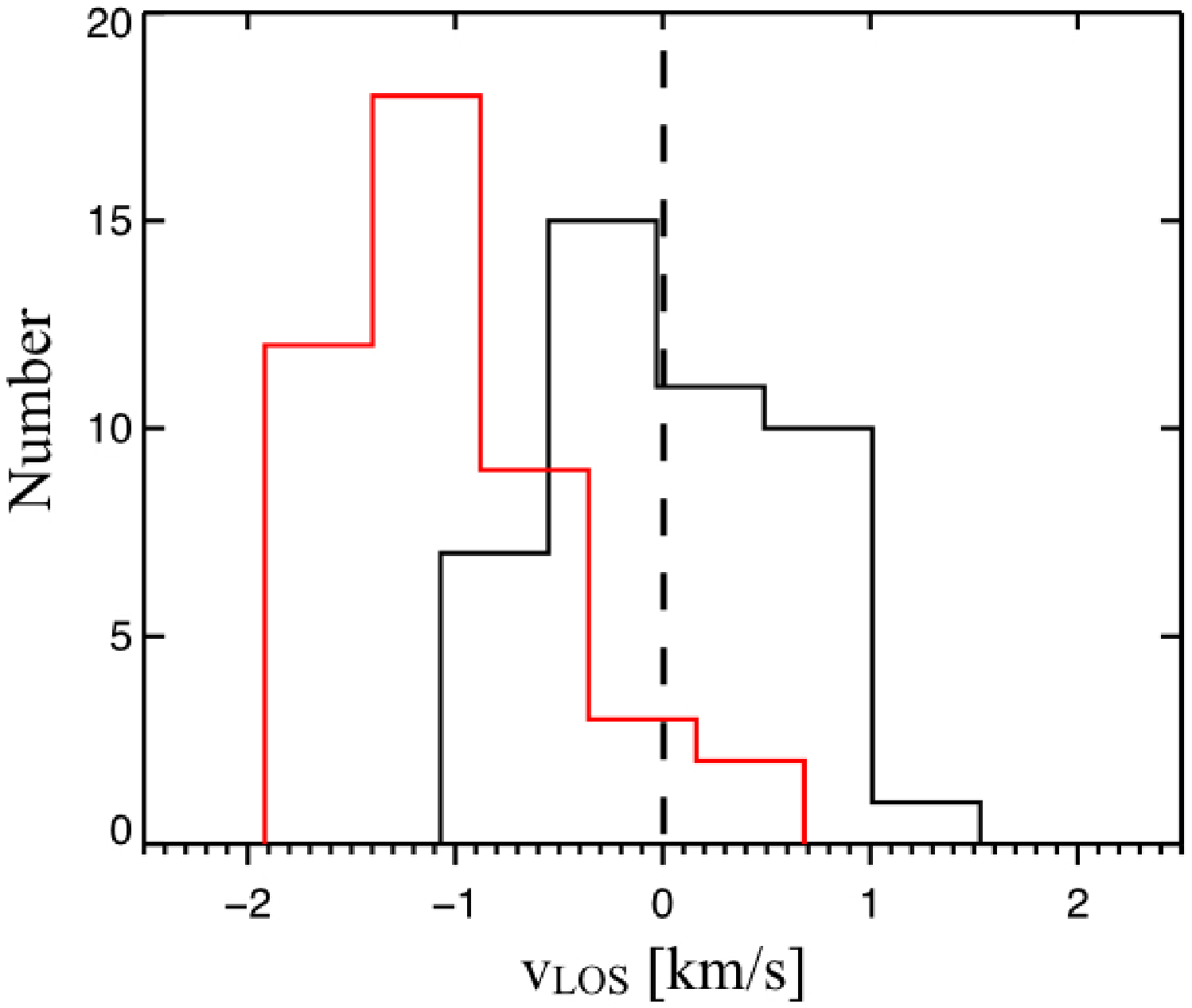}}(c) &
    {\includegraphics[width=0.4\textwidth]{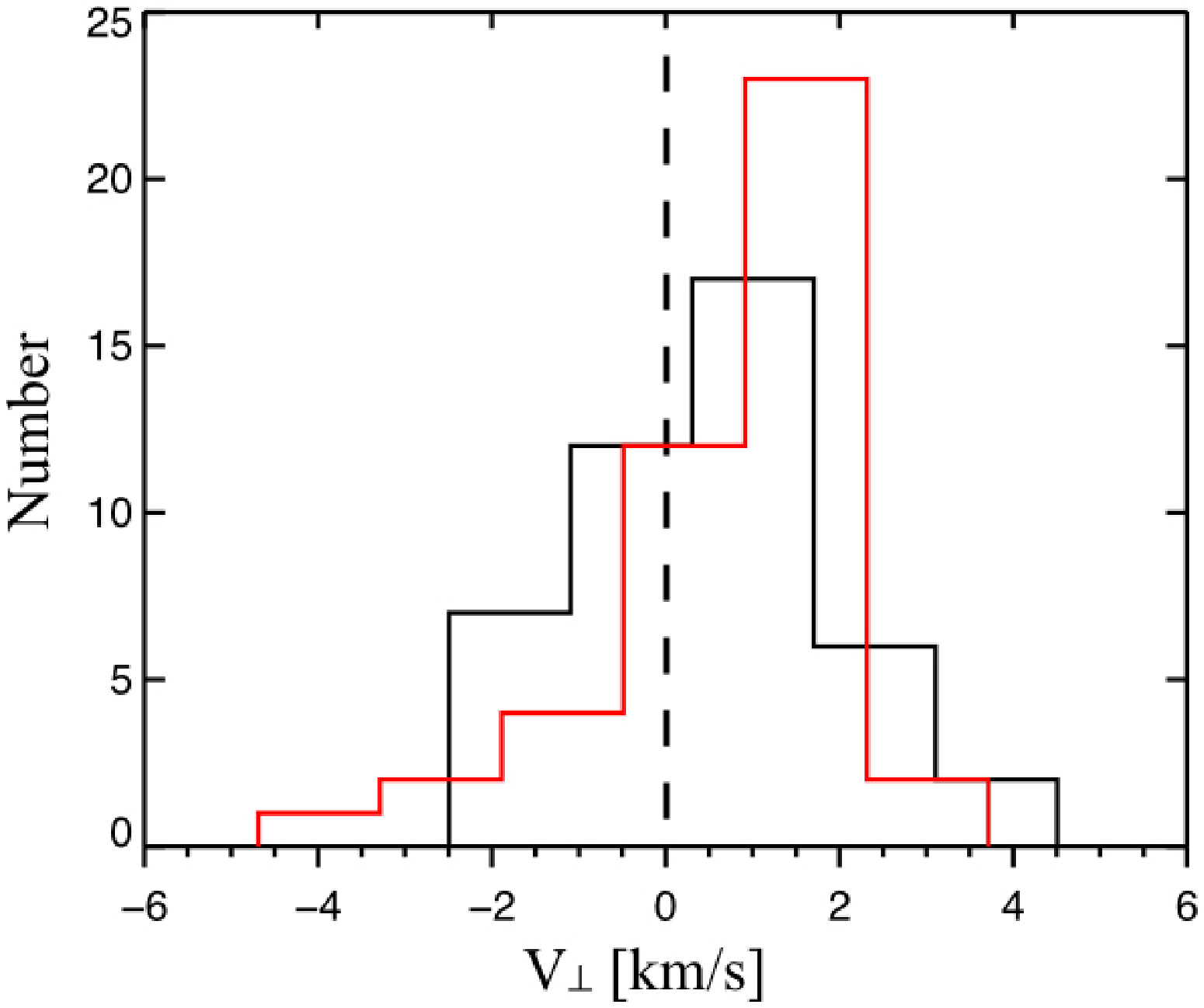}}(d)\\
    
    {\includegraphics[width=0.4\textwidth]{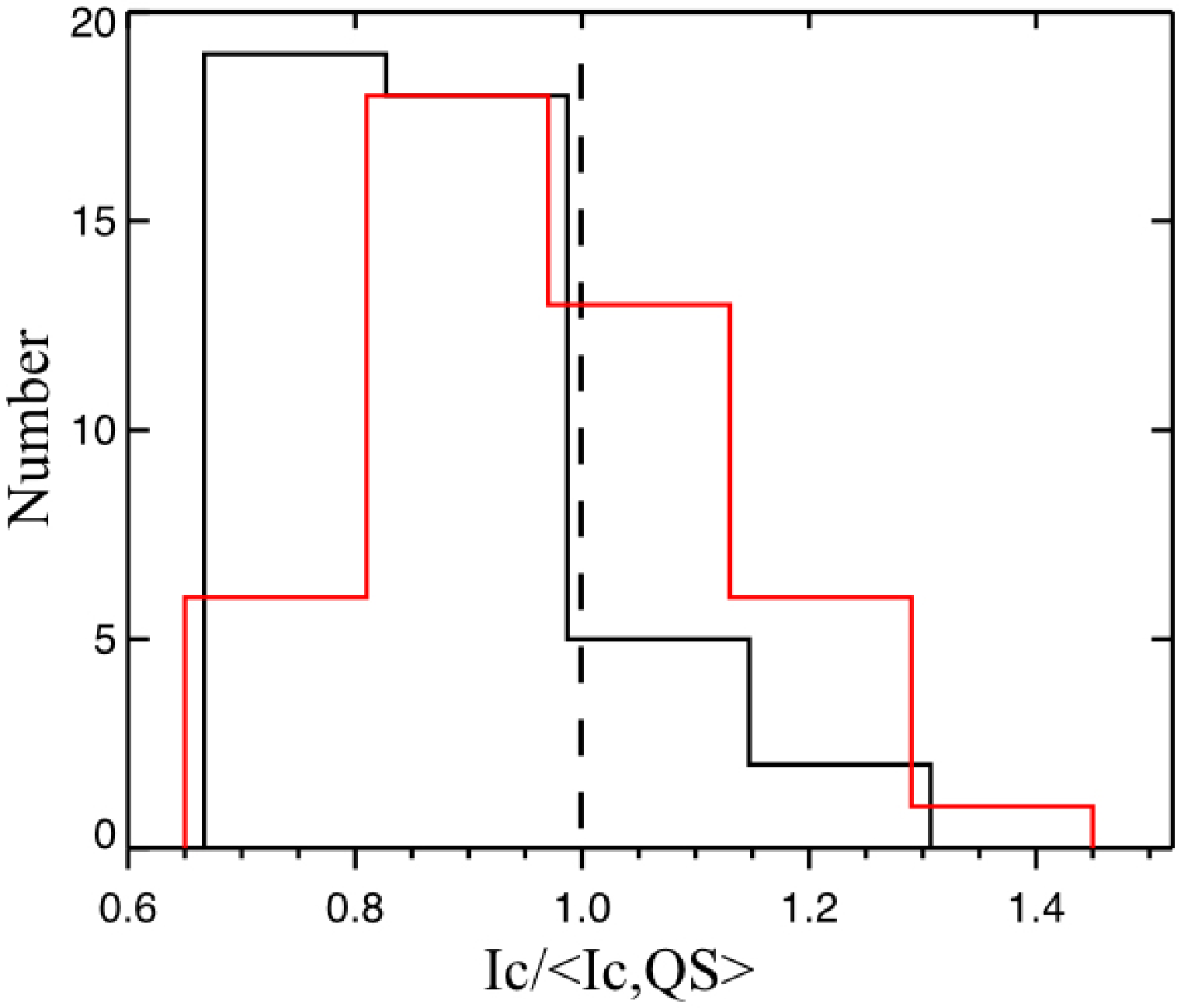}}(e)&
    {\includegraphics[width=0.4\textwidth]{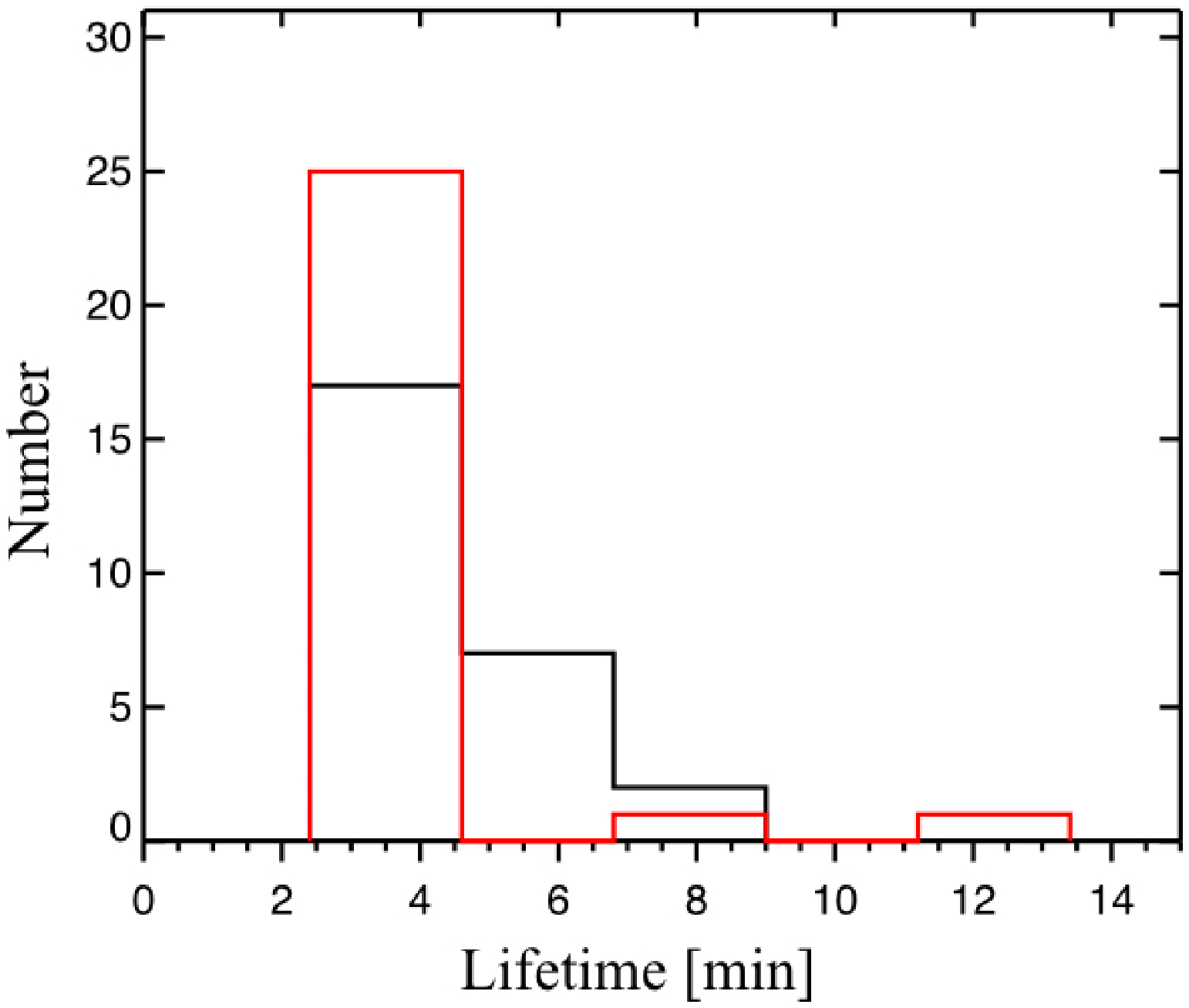}}(f)   
\end{array}$
\caption{Histograms of the lifetime average of $B$ (a), $\gamma$ (b), $v_{LOS}$ (c), $V_\perp$ (d) (see the right panel of Figure \ref{fig:a} for definition),  at the magnetic centroid position (e) and, lifetime of MMFs (f). Red (black) plots represent opposite- (same-) polarity MMFs. In panel (c) negative Doppler velocities correspond to blueshifts. In panel (d) positive velocities represent MMFs moving away from the pore border and negative velocities represent MMFs moving towards the pore.\label{fig:d}}
\centering
\end{figure*}

(1) For MMFs of both polarities, the intrinsic field strength (Figure \ref{fig:d} a) displays a broad distribution: for opposite-polarity MMFs it is in the range 250 -- 1300 G, and for the same-polarity MMFs it is in the range 250 -- 900 G. We also find that, as the distance from the pore border increases the field strength decreases for MMFs of both types (correlation coefficients are $-0.60$ and $-0.50$ respectively, for opposite- and same-polarity MMFs). However, the field strength displays random fluctuations with time for MMFs of both polarities.

(2) The histograms of field inclinations (Figure \ref{fig:d} b) show that MMFs of both polarities are highly inclined. The mean inclination of opposite polarity MMFs is about $75^\circ$ and that of same-polarity MMFs is about $106^\circ$. With increasing distance from the pore boundary the magnetic field displays a tendency to become more vertical for MMFs of opposite polarity (correlation coefficient: $-0.46$) whereas for the same-polarity MMFs this trend is comparatively weak. These observations are in line with the study on unipolar MMFs around a pore by Criscuoli et al (2012), who showed that field vectors of MMFs can be highly inclined, although a vertical orientation is the most probable configuration. They also reported that, for both types of the MMFs the field becomes more vertical with increasing distance from the pore border.

(3) The distributions of $v_{LOS}$ (Figure \ref{fig:d} c) show that MMFs of opposite polarity are characterized by preferential upflows. The peak of the distribution is around 1 km s$^{-1}$ and the mean value is $\sim$1.1 km s$^{-1}$. In contrast, the same-polarity MMFs are on average unshifted. Opposite-polarity MMFs exhibit a slow decrease in upflow as their distance from the pore border increases (correlation coefficient: 0.68). Some MMFs of both polarities show both up- and downflows during their lifetime.  

This result agrees partly with that of \cite{ser}. The authors had shown that opposite-polarity MMFs are associated with upflows and same-polarity MMFs with downflows. Also they found, the upflows associated with the opposite-polarity MMFs decrease slowly with distance from the pore border.

(4) Panel (d) of Figure \ref{fig:d} presents the histograms of the perpendicular component of the proper motion velocity ($V_\perp$) of the MMFs. MMFs of opposite polarity have speeds in the range [$-4.7$ -- 3.4] km s$^{-1}$. They move away from the pore border with a mean speed of $\sim$ 1.2 km s$^{-1}$. The mean speed of these MMFs is $\sim$ 1.5 km s$^{-1}$ when they were first detected and is $\sim$ 1.1 km s$^{-1}$ when they were recognized for the last time. The mean speed of the opposite-polarity MMFs which move toward the pore is $\sim$ $-2.0$ km s$^{-1}$. 

MMFs of same polarity have $V_\perp$ in the range of [$-2.5$ -- 3.7] km s$^{-1}$. They move away from the pore with a mean speed of 1.3 km s$^{-1}$. The mean value of $V_\perp$ corresponding to the first and last detection of these MMFs is $\sim$ 2.2 and 1.2 km s$^{-1}$, respectively. The same polarity MMFs which move toward the pore have a mean speed of $\sim$ $-1.1$ km s$^{-1}$.

The mean initial speed with which the MMFs of both polarities move away from the pore is close to the value of 1.8 km s$^{-1}$ found by \cite{hag} using MDI data in the high resolution mode. The mean $V_\perp$ value of the MMFs that move away is higher than the value of 0.45 km s$^{-1}$ reported by \cite{zh1} and of 0.34 km s$^{-1}$ obtained by \cite{kuboa} for MMFs around sunspots, both using MDI data. For bipolar MMFs observed around a sunspot without penumbra (using Hinode data), \cite{zu} found that the MMFs have a typical horizontal velocity of 0.7 km s$^{-1}$. 

For MMFs of opposite polarity $V_\parallel$ is in the range [$-4.2$ -- 3.4] km s$^{-1}$ with a mean value of $-0.17$ km s$^{-1}$. $V_\parallel$ of same-polarity MMFs also shows a broad range  from $-2.8$ to 3.9 km s$^{-1}$, while the average value of the distribution is 0.12 km s$^{-1}$. 



(5) Figure \ref{fig:d} (e) shows that the majority of the same-polarity MMFs stays below the mean quiet-Sun intensity level. We find that the distribution has a broad range with a mean value of 0.86. The opposite-polarity MMFs also exhibits a broad distribution ranging from 30$\%$ below- to 40$\%$ above the average quiet-Sun intensity, and mean value of 0.97. 

We find that for MMFs of both polarities the intensity varies randomly with time. For most of the same-polarity MMFs, the instantaneous intensity stays below the mean quiet-Sun value throughout their lifetime. In the case of opposite-polarity MMFs the intensity is widely distributed over time.

(6) Histograms of the lifetime of the MMFs are presented in Figure \ref{fig:d} (f). For these plots we consider only those MMFs which were born and disappeared during the observation. Irrespective of the polarity, most of the MMFs have a lifetime of less than 5 minutes. The observed average lifetime of the same-polarity MMFs is about 4.1 minutes and that of the opposite polarity MMFs is 3.6 minutes.


\section{Summary and Discussion} \label{sec:conc} 

We investigated the physical properties of moving magnetic features observed outside a pore using \sunrise{}II/IMaX data. Our main results obtained from the statistical analysis are summarized as follows: MMFs of the same and opposite polarity to that of the pore are observed at one side of the pore boundary. Most of these MMFs are of sub-arcsecond size. The majority of the MMFs move away from the pore border with a preference for the radial direction. Opposite-polarity MMFs are generally blue shifted and same-polarity MMFs are on average unshifted. MMFs of both polarities are characterized by weak and inclined magnetic fields.

Although the motion of the MMFs we observed are greatly influenced by the granulation they exhibit a clear preference for a radial outward motion. The tracks followed by the MMFs bear a resemblance to the trajectories of intergranular magnetic bright points (MBPs) reported by \cite{sah} using \sunrise{}/SuFI data \citep{gan}. The authors demonstrated that the trajectories of the MBPs result from the superposition of random motions caused by granular evolution and intergranular turbulence and systematic motions caused by steady granular evolution and mesogranular and supergranular flows. They also found that these MBPs are super-diffusive in nature. It would be worth investigating whether the MMFs show a similar dispersion trend. 

We found that MMFs of both polarities are characterized by inclined magnetic fields, with most of them having field strengths below 1 kG. A study, based on ASP data by \cite{kuboa} on MMFs around a sunspot had shown that many MMFs that are located on lines extrapolated from the horizontal component of the penumbra are associated with inclined magnetic fields with field strength below 1 kG. \cite{ser} found that the MMFs of both polarities observed around a pore have a wide distribution of magnetic field strengths ranging from 500 -- 1700 G. The authors mentioned that the most probable value of $B$ is above 1 kG and that the magnetic fields are most probably vertically oriented. The physical parameters of the MMFs obtained in this study are compared with those found by some other studies in Table \ref{tab:t1}.

MMFs of both polarities are found to be associated with inclined fields. However, with increasing distance from the pore border they show a trend of becoming less inclined. If these are individual MMFs (i.e. not members of pairs), then it may have to do with the strong magnetic canopy of the pore, which forces the magnetic field of the MMFs to remain relatively horizontal. With increasing distance from the pore boundary, the canopy lies higher and the canopy field is weaker, so that once they are far enough from the pore field they become more vertical. 

Our ROI is special when compared to other areas around the pore in that opposite-polarity MMFs prevail in this region. Figure \ref{fig:d} (c) shows that opposite-polarity MMFs are preferentially associated with upflows. This picture is supported by the fact that the opposite-polarity MMFs are on average brighter than the same-polarity MMFs, although only around half the opposite-polarity MMFs are brighter than the average quiet-Sun. However, with increasing distance from the pore these MMFs exhibit a slow decrease in blueshift, which implies that they possibly end up in the intergranular region just as the same-polarity MMFs.

The majority of the same-polarity MMFs is characterized by intensity values lower than the average quiet-Sun value. This suggests that most of these MMFs were born and remained in the intergranular lanes. This is similar to the picture presented by \cite{cam}. Using MHD simulations of a pore the authors had shown that intergranular lanes are preferentially occupied by features of the same polarity as the pore. This could be one reason why opposite-polarity MMFs are shorter-lived than the same polarity ones, because as the opposite-polarity MMFs move into the intergranular lanes the chances are high that their magnetic flux is eventually cancelled out by the existing same-polarity flux. This could provide an explanation as to why same-polarity MMFs dominate the region beyond 2 Mm from the pore border.  


\begin{acknowledgements}
A.J.K thanks R. Cameron and H. Uitenbroek for helpful comments and suggestions. The German contribution to \sunrise{} and its reflight was funded by the Max Planck Foundation, the Strategic Innovations Fund of the President of the Max Planck Society (MPG), DLR, and private donations by supporting members of the Max Planck Society, which is gratefully acknowledged. The Spanish contribution was funded by the Ministerio de Econom\'i­a y Competitividad under Projects ESP2013-47349-C6 and ESP2014-56169-C6, partially using European FEDER funds. The HAO contribution was partly funded through NASA grant number NNX13AE95G. This work was partly supported by the BK21 plus program through the National Research Foundation (NRF) funded by the Ministry of Education of Korea. 
\end{acknowledgements}

\floattable
\begin{deluxetable*}{ccccccc}
\tabletypesize{\scriptsize}
\tablecaption{ Comparison of the physical parameters of MMFs obtained in this study with
some of those in the literature.\label{tab:t1}}
\tablecolumns{7}
\tablewidth{0pt}
\tablehead{
\colhead{Reference} & \colhead{Instrument} & \colhead{Mean Flux} & \colhead{$B$} & \colhead{$\gamma$} & \colhead{$v_{LOS}$} & \colhead{Mean horizontal velocity\tablenotemark{a}}\\
\colhead{} & \colhead{} & \colhead{(Mx)} & \colhead{(G)} & \colhead{(deg.)} & \colhead{(km s$^{-1}$)} & \colhead{(km s$^{-1}$)}}
\startdata
This study     & $\sunrise{}$II/IMaX  & 1.16$\times$10$^{17}$  & 250 -- 1300       & 55 -- 85          & upflow                & 1.2\\
                     &                                 & 1.19$\times$10$^{17}$   & 250 -- 900         & 97 --  128       & no preference    & $"$\\               
\cite{ser}       &   DST/IBIS               &        ...                             & 500 -- 1700       & 10 -- 50          & upflow                & $\sim$ 1\\
                     &                                 &        ...     		               &        $"$	           & 120 -- 160      & downflow           &   ...\\
\cite{kuboa}  &  ASP and MDI         &        ...       		      & 300 -- 1600        & 40 -- 70	   &     ...                   & 0.34\\
                     &                                &        ...    		               &      $"$ 	           & 90 -- 170        &     ...                    &   $"$\\
\cite{zh1}     &  MDI                         & 3.6$\times$10$^{18}$     &     ...	           &	...		   &     ...                   & 0.45\\
\enddata
\tablenotetext{a}{For our study mean value of $V_\perp$ (perpendicular component of proper motion velocity) of MMFs moving outward is given.}
\tablecomments{For each study except \cite{kuboa}, parameters in the first line are for opposite-polarity MMFs and those in the second line are for same-polarity MMFs. For the study by \cite{kuboa} the first line represents values for same-polarity MMFs.}
\end{deluxetable*}

\newpage

\bibliography{bib}

\end{document}